\documentclass{ws-ijmpe}
\usepackage[super,compress]{cite}
\usepackage{graphicx}

\newcommand{\al}{$\alpha$}

\newcommand{\raa}{($\alpha$,$\alpha$)}

\newcommand{\rag}{($\alpha$,$\gamma$)}
\newcommand{\ran}{($\alpha$,n)}

\newcommand{\rap}{($\alpha$,p)}

\newcommand{\rpg}{(p,$\gamma$)}

\newcommand{\stot}{$\sigma_{\rm{reac}}$}

\newcommand{\Nsv}{$N_A$$\left< \sigma v \right>$}

\newcommand{\canull}{$^{40}$Ca}

\newcommand{\sciii}{$^{43}$Sc}

\newcommand{\tiiii}{$^{43}$Ti}
\newcommand{\tiiv}{$^{44}$Ti}

\newcommand{\KS}{K\&S}
\newcommand{\WSsq}{WS$^2$}

\begin{document}

\markboth{Peter Mohr}{Cross sections at sub-Coulomb energies}

\catchline{}{}{}{}{}

\title{
  Cross sections at sub-Coulomb energies:\\
  full optical model vs.\ barrier transmission for $^{40}$Ca + $\alpha$
  }

\author{
  Peter Mohr
  }

\address{
  Institute for Nuclear Research (Atomki),
  P.O. Box 51, Debrecen H-4001, Hungary \\
  Diakonie-Klinikum, Schw\"{a}bisch Hall D-74523, Germany \\
mohr@atomki.mta.hu}

\maketitle

\begin{history}
\received{Day Month Year}
\revised{Day Month Year}
\end{history}

\begin{abstract}
Cross sections for $^{40}$Ca + $\alpha$ at low energies have been calculated
from two different models and three different $\alpha$-nucleus potentials. The
first model determines the cross sections from the barrier transmission in a
real nuclear potential. Second, cross sections are derived within the optical
model using a complex nuclear potential. The excitation functions from barrier
transmission are smooth whereas the excitation functions from the optical
model show a significant sensitivity to the chosen imaginary potential. Cross
sections far below the Coulomb barrier are lower from barrier transmission
than from the optical model. This difference is explained by additional
absorption in the tail of the imaginary part of the potential in the optical
model. At higher energies the calculations from the two models and all
$\alpha$-nucleus potentials converge. Finally, in contradiction to another
recent study where a double-folding potential failed in a WKB calculation, the
applicability of double-folding potentials for $^{40}$Ca + $\alpha$ at low
energies is clearly confirmed in the present analysis for the simple barrier
transmission model and for the full optical model calculation.
\end{abstract}




\section{Introduction}
\label{sec:intro}
Cross sections at low energies around and below the Coulomb barrier play an
important role in various areas of nuclear physics. Because of the high
abundances of hydrogen and helium in stars, proton- and \al -induced reactions
play a fundamental role in nuclear astrophysics
\cite{Rolfs_Book,Iliadis_Book}. Fusion reactions between heavier nuclei are
essential to extend the chart of nuclei e.g.\ towards superheavy nuclei
\cite{Oganessian_RPP2015_superheavy}. In general, the calculation of cross
sections is based on the two-body Schr\"odinger equation which in turn
requires a well chosen potential $U(r)$ between the colliding nuclei. This
potential $U(r)$ is composed of a nuclear part $V_N(r)$ and a Coulomb part
$V_C(r)$:
\begin{equation}
  U(r) = V_N(r) + V_C(r) \quad \quad .
  \label{eq:pot}
\end{equation}
Two different models are investigated in this study. In a first model, only
the real part of the nuclear potential is considered to determine transmission
coefficients. In a second model, the optical model, cross sections are
calculated from a complex nuclear potential. The models will be presented in
Sec.~\ref{sec:mod}, the potentials will be discussed in Sec.~\ref{sec:pot},
and the results will be shown in Sec.~\ref{sec:res}. Advantages and
disadvantages of the two approaches will be analyzed in Sec.~\ref{sec:disc},
and the findings will be summarized in Sec.~\ref{sec:summ}.

This investigation was triggered by a recent publication by Koyuncu and Soylu
in this journal \cite{Koyuncu_IJMPE2017} (hereafter: \KS ). In that work the
fusion of \al\ + \canull\ was investigated using a semi-classical
approximation of the first model of this study in combination with several \al
-nucleus potentials. The scope of the present study is threefold: First, the
work of \KS\ is extended by optical model calculations to obtain a better
understanding of the energy dependence of low-energy cross sections. Second, a
strong statement has been made by \KS\ that the ``DF potential $\ldots$ has
failed to produce \al\ + \canull\ cross sections at low energies''; the
present work attempts to verify or disprove this claim. From earlier
investigations \cite{Atzrott_PRC1993_alpha,Ohkubo_PRC1988_44ti} of \canull
\raa \canull\ elastic scattering and bound state properties of \tiiv\ =
\canull\ $\otimes$ \al\ it is expected that double-folding potentials are well
suited for \canull\ + \al\ at low energies, and it was concluded that
potentials with similar shapes (but various parametrizations) can be applied
successfully to \canull\ + \al\ \cite{Michel_PRC1986_44ti}. For further
information on folding potentials for \canull\ + \al , see also
\cite{Khoa_JPG2007_rainbow,Gils_NPA1987_scat,Khoa_PRC1997_folding,Ohkubo_PRC2016_cluster}. Third,
the relevance of the imaginary part on the calculated low-energy cross
sections in the optical model is investigated, and a strong sensitivity to the
tail of the imaginary potential is found for energies far below the Coulomb
barrier.

All energies will be given in the center-of-mass (cm) system throughout this
paper (except explicitly noted).

\section{Calculations and results}
\label{sec:calc}
\subsection{Models}
\label{sec:mod}
Two different models in combination with three different \al -nucleus
potentials were used to calculate the cross sections of \al\ + \canull\ at low
energies around and below the Coulomb barrier. Slightly depending on the
chosen nuclear potential, the effective barrier is located at radii of $r
\approx 8$ fm and has a height of $6.5 - 6.8$ MeV. 

The total (non-elastic) reaction cross section \stot\ in both models results
from the sum over the partial cross sections $\sigma_L$ for each contributing
partial wave with angular momentum $L$:
\begin{equation}
  \sigma_{\rm{reac}} = \sum_L \sigma_L \quad \quad .
  \label{eq:partial}
\end{equation}
The partial $\sigma_L$ are calculated by solving the Schr\"odinger equation
for angular momenta $L$ from $L = 0$ to $L_{\rm{max}}$. In the energy range
under study ($E \le 10$ MeV), a maximum angular momentum of $L_{\rm{max}}
\approx 10$ is sufficient for the calculation of \stot\ (see also
Sec.~\ref{sec:res}). For finite angular momenta $L > 0$, the potential $U(r)$
in Eq.~(\ref{eq:pot}) in the Schr\"odinger equation has to be complemented by
the usual centrifugal potential
\begin{equation}
  V_L(r) = \frac{L (L+1) \hbar^2}{2 \mu r^2}
  \label{eq:pot_cent}
\end{equation}
with the reduced mass $\mu$ of the system under study.

The Coulomb potential in Eq.~(\ref{eq:pot}) is calculated from the model of a
homogeneously charged sphere. The chosen Coulomb radius $R_C$ will be given
for each potential under study in Sec.~\ref{sec:pot}.

\subsubsection{Model 1: pure barrier transmission in a real potential}
\label{sec:mod1}
The first model uses a real potential to calculate the barrier transmission
$T_L$ through the Coulomb (plus centrifugal) barrier. This leads to the
reaction cross section
\begin{equation}
  \sigma_{\rm{reac}} = \sum_L \sigma_L =
  \frac{\pi}{k^2} \sum_L \, (2L+1) \, T_L(E)
\label{eq:BTM}
\end{equation}
where the wave number $k$ is related to the energy by $E = \frac{\hbar^2
  k^2}{2\mu}$. This model is widely used for the calculation of fusion cross
sections (see e.g.\ \KS\ and references therein). Complications with the
calculation of the Coulomb functions at very low energies can be avoided in
this model by changing to a semi-classical treatment; such a semi-classical
treatment, the so-called WKB method, was used by \KS . Results from the pure
barrier transmission model will be labeled by ``pBTM'' in the
following. Calculations for the pBTM have been made using the CCFULL code
\cite{Hagino_CPC1999_ccfull}. As the CCFULL code is based on Woods-Saxon
potentials only, the input routine of this code had to be adapted to read the
double-folding potentials from an external numerical file. The CCFULL code
solves the coupled-channel equations, as given in Eq.~(1) of
\cite{Hagino_CPC1999_ccfull}. In the present case, the explicit couplings by
the matrix elements $V_{nm}$ in Eq.~(1) of \cite{Hagino_CPC1999_ccfull} to
inelastic channels (i.e., excited states in \canull ) were switched
off. Technically, CCFULL applies the so-called modified Numerov method to
solve the coupled-channel equations numerically from a minimum radius
$r_{\rm{min}}$ (calculated from the minimum position of the Coulomb pocket
inside the barrier) to a maximum radius $r_{\rm{max}}$ outside the barrier
(where the nuclear potential becomes negligible); at $r_{\rm{max}}$ the
numerically integrated wave function is matched to the Coulomb wave
function. The transmissions $T_L$ are calculated from the amplitude of the
wave function, see Eqs.~(11), (16) and (17) in \cite{Hagino_CPC1999_ccfull}.
In the semi-classical WKB method, $r_{\rm{min}}$ and $r_{\rm{max}}$ are simply
taken as the classical turning points where the total potential $V(r) = V_N(r)
+ V_C(r) + V_L(r)$ is identical to the energy $E$. Note that at low energies
$r_{\rm{max}}$ in the WKB approximation is typically much larger than in the
CCFULL calculation, whereas $r_{\rm{min}}$ is similar in both approaches.

\subsubsection{Model 2: optical model}
\label{sec:mod2}
The second model applies a complex nuclear potential $V_N(r) = V_R(r) +
iW(r)$. Because of its similarities to optics, this model is generally called
optical model (labeled ``OM''). Now the total (non-elastic) reaction cross
section \stot\ results from the following equation:
\begin{equation}
  \sigma_{\rm{reac}} = \sum_L \sigma_L =
  \frac{\pi}{k^2} \sum_L \, (2L+1) \, \bigl[1 - \eta_L^2(E) \bigr]
  \quad \quad .
\label{eq:OM}
\end{equation}
The $\eta_L$ are the real reflexion coefficients which result from the
solution of the Schr\"odinger equation using the complex nuclear potential
$V_N(r)$. The real $\eta_L$ and the real phase shifts $\delta_L$ are related
to the complex scattering matrix elements $S_L$ by
\begin{equation}
  S_L = \eta_L \, \exp{(2 i \delta_L)} \quad \quad .
  \label{eq:SL}
\end{equation}
Formally, Eq.~(\ref{eq:OM}) is identical to the previous Eq.~(\ref{eq:BTM})
for the pBTM because the $(1 - \eta_L^2)$ in Eq.~(\ref{eq:OM}) are also called
transmissions. However, there is also an essential difference: for a real
nuclear potential (as used in the pBTM), one finds pure elastic scattering
with $\eta_L = 1$ in the OM; the resulting phase shifts $\delta_L \ne 0$
reflect the real potential $V(r)$ and define the elastic scattering angular
distribution. According to Eq.~(\ref{eq:OM}), $\eta_L = 1$ for all $L$ leads
to vanishing partial reaction cross sections $\sigma_L = 0$ and thus \stot\ $ =
0$. For any real nuclear potential without imaginary part, the flux of
incoming particles completely remains in the elastic channel. Finite reaction
cross sections $\sigma_L$ in the OM finally result from the imaginary part
$W(r)$ of the nuclear potential $V_N(r)$. This aspect will be discussed
further in Sec.~\ref{sec:comp}.

The OM is very widely used in nuclear physics for the analysis of elastic
scattering and total reaction cross sections. Furthermore, the OM is the basic
building block of statistical model (SM) calculations where the formation
cross section of a compound nucleus is taken from the total reaction cross
section \stot . Cross sections of individual exit channels in the SM are also
calculated from optical potentials in the respective particle exit channels
(and from the $\gamma$-ray strength function for the capture channel)
\cite{Hauser_PR1952,Rauscher_IJMPE2011}.

OM calculations have been performed using the code a0 \cite{A0}; the TALYS
code \cite{TALYS-V19} was applied for additional SM calculations. These
calculations can be compared to experimental data for the \canull \rap
\sciii\ reaction \cite{Howard_APJ1974_a-X}.

\subsubsection{Comparison of the models}
\label{sec:comp}
There is one essential difference between the pBTM and the OM: The pBTM
provides the transmissions $T_L$ in Eq.~(\ref{eq:BTM}) in a real potential,
and by definition it is assumed that fusion occurs as soon as the incoming \al
-particle has tunneled through the Coulomb barrier. Contrary, in the OM an
imaginary part of the potential is required to describe absorption and to
remove flux of the incoming \al -particles from the elastic channel to
non-elastic channels. In a simplified view, also in the OM the incoming \al
-particle has to tunnel through the Coulomb barrier; this tunneling is similar
to the pBTM. However, absorption in the OM can only occur if the \al -particle
``feels'' the imaginary part $W(r)$ at smaller radii in the surface and
interior of the compound nucleus.  As a consequence, the total reaction cross
section in the OM results from the solution of the time-independent
Schr\"odinger equation and depends on both, the real part $V_R(r)$ and the
imaginary part $W(r)$, of the nuclear potential.

For completeness it has to be mentioned that the semi-classical WKB method is
simply an approximation of the pBTM. In the WKB approximation, the $T_L$ in
Eq.~(\ref{eq:BTM}) are calculated from the damping of the wave function in the
barrier between the classical turning points according to Eqs.~(3) and (4) in
\KS .  The WKB approximation simplifies the calculation of the $T_L$
especially at very low energies because a calculation of the Coulomb wave
functions is not necessary. The disadvantage of the WKB calculation is the
well-known sensitivity to the turning points. The WKB method is widely used in
\al -decay studies. A comparison between the semi-classical WKB approximation
and a fully quantum-mechanical calculation of \al -decay half-lives was
already given in earlier work for $^{212}$Po $\rightarrow$ $^{208}$Pb +
\al\ \cite{Mohr_PRC2006_superheavy} and $^{104}$Te $\rightarrow$ $^{100}$Sn +
\al\ \cite{Mohr_EPJA2007_A100}, and only relatively small deviations below
30\% were found in all cases.

It is an interesting question whether the same real part of the potential
should be used in the pBTM and in the OM calculations. The role of the real
part is similar in both models as it describes the tunneling through the
barrier. When the same real potentials are used in the pBTM and in the OM,
indeed similar total cross sections are found in both models at energies above
and slightly below the barrier. However, at energies far below the barrier the
cross sections in the OM are significantly higher than the cross sections in
the pBTM. This finding is related to the properties of the imaginary part of
the potential in the OM calculations and will be explained further in
Sec.~\ref{sec:disc}.

From the above general remarks it is obvious that the OM is more
microscopically founded than the simpler pBTM; this is a clear advantage of
the OM. But at the same time, the shape of the imaginary potential $W(r)$ has
to be well-known for the reliable prediction of cross sections in the OM,
especially at low energies. Unfortunately it is not a simple task to fix the
imaginary potential $W(r)$ (e.g., from the analysis of elastic scattering
angular distributions). Hence, the parametrization of the imaginary potential
$W(r)$ leads to significant uncertainties for the prediction of cross sections
at very low energies in the OM (see also Sec.~\ref{sec:disc}).

The advantage of the simpler pBTM is the lower number of adjustable
parameters. The real part of the nuclear potential is relatively well
constrained (e.g., by the folding procedure), and thus the uncertainties from
the choice of parameters are relatively small in the pBTM. Furthermore, at
energies far below the Coulomb barrier, the calculation of the Coulomb
functions becomes numerically complicated. At these low energies the pBTM can
easily switch to semi-classical approximations like the WKB method which are
widely used e.g.\ in the calculation of \al -decay half-lives.

\subsection{Potentials}
\label{sec:pot}
The basic ingredient for the following calculations is the \al -nucleus
potential. Several options have already been chosen by \KS , and it was shown
that the calculated cross sections \stot\ are close to each other with the
exception of the double-folding potential which showed a by far flatter energy
dependence in the excitation function. This leads to dramatically higher cross
sections by many orders of magnitude at very low energies (see Fig.~1 in \KS
). However, as soon as the double-folding potential was fitted by a squared
Woods-Saxon potential (\WSsq ), the calculated excitation function of
\KS\ remained regular.

Therefore, in the following I focus on three potentials:
\begin{itemlist}
 \item The \WSsq\ potential of \KS\ with $V_0 = -270$ MeV, $R = 4.35$ fm, and
   $a = 1.26$ fm which was fitted by \KS\ to their double-folding potential.
 \item The ATOMKI-V1 potential which uses a double-folding potential in the
   real part and a Woods-Saxon parametrization of surface type in the
   imaginary part \cite{Mohr_ADNDT2013_atomki-v1}. Here the nuclear densities
   of \canull\ and \al\ were derived from experimental charge density
   distributions \cite{deVries_ADNDT1987}. To be specific, for \canull\ the
   Fourier-Bessel parameterization of \cite{Emrich_NPA1983_ca40-ca48} and for
   \al\ a sum of Gaussians from \cite{Sick_PLB1982} was used; the underlying
   electron scattering data cover the largest range of momentum transfers for
   the chosen density distributions. The interaction was calculated at an
   average energy $E_{\alpha,{\rm{lab}}} = 5$ MeV from the density-dependent
   M3Y parameters as listed in Table 1 of \cite{Mohr_ADNDT2013_atomki-v1}; for
   further details, see also
   \cite{Atzrott_PRC1993_alpha,Kobos_NPA1984_folding,Satchler_NPA1979_folding}.
   The parameters of the ATOMKI-V1 potential, namely the strength parameter
   $\lambda$ of the real part and the depth, radius, and diffuseness of the
   imaginary part of surface Woods-Saxon type, were adjusted to elastic
   \al\ scattering data in the $89 \le A \le 144$ mass range at low energies.
 \item The well-established simple 4-parameter Woods-Saxon potential by
   McFadden and Satchler \cite{McFadden_NPA1966_aomp} (McF) with $V_0 = -185$
   MeV, $W_0 = -25$ MeV, $R_0 = 1.4$ fm (to be multiplied by $A_T^{1/3}$), and
   $a = 0.52$ fm which is known to work very well in this mass region
   \cite{Mohr_EPJA2015_A20-50,Mohr_PRC2018_38ar}. The parameters of the McF
   potential were derived from elastic \al\ scattering at
   $E_{\alpha,{\rm{lab}}} = 25$ MeV for a wide range of masses of the target
   nuclei.
\end{itemlist}

The Coulomb potential $V_C$ in Eq.~(\ref{eq:pot}) is calculated from a
homogeneously charged sphere with a reduced Coulomb radius $R_{C,0} = 1.3$ fm
for the McF and the \WSsq\ potentials (to be multiplied by $A_T^{1/3}$), and
$R_C$ was set to the root-mean-square radius of the folding potential for the
ATOMKI-V1 potential: $R_C = 4.231$ fm ($R_{C,0} = 1.237$ fm).

The minor difference in the choice of $R_C$ for the different potentials under
study does not lead to significant variations in the calculated cross
sections. E.g., varying the Coulomb radii $R_{C,0}$ in a wider range between
1.1 fm and 1.5 fm (corresponding to $R_C = 3.76$ fm to 5.13 fm) changes the
calculated cross sections \stot\ by less than 1\% over the whole energy range
of the present study; this was tested in the pBTM in combination with the real
part of the McF potential and in the OM in combination with the full McF
potential. Major changes of \stot\ would only be obtained for much larger
Coulomb radii $R_C \approx 8$ fm around the effective barrier. As long as the
Coulomb radius $R_C$ is much smaller, the Coulomb potential $V_C(r)$ outside
$R_C$ shows the same $1/r$ behavior in the barrier and thus does practically
not affect the effective barrier and the calculated cross sections \stot .

The ATOMKI-V1 potential was selected because it is based on a double-folding
procedure and can be nicely compared to the WS$^2$ potential by \KS . The McF
potential was chosen because cross sections of \al -induced reactions are well
reproduced in the mass region $20 \le A \le 50$, see e.g.\
Refs.~\cite{Mohr_EPJA2015_A20-50,Mohr_PRC2018_38ar}.

For completeness, further tests have been made using the global potentials by
Demetriou {\it et al.}\ \cite{Demetriou_NPA2002_aomp} and Avrigeanu {\it et
  al.}\ \cite{Avrigeanu_PRC2014_aomp} which have been determined to calculate
low-energy cross sections of \al -induced reactions. Both potentials show
similar cross sections as the McF potential above 5 MeV and slightly lower
cross sections at lower energies. However, the deviation remains within a
factor of two for the Avrigeanu potential for the full energy range under
study and within a factor of three for the Demetriou potential for almost the
full energy range (except the lowest energies below 3 MeV where a discrepancy
of one order of magnitude is reached). Finally, a specially shaped (1 +
Gaussian) $\times$ (WS + WS$^3$) potential was successfully applied for the
desciption of \al -cluster states over a wide mass range, including the
example of \tiiv\ = \canull\ + \al\ \cite{Souza_PLB2019_cluster}. The shape of
the barrier of this potential is practically identical to the ATOMKI-V1
potential, and consequently the resulting cross sections in the pBTM do not
deviate by more than 15\% in the energy range under study.

\begin{figure}[th]
\centerline{\includegraphics[width=0.75\columnwidth]{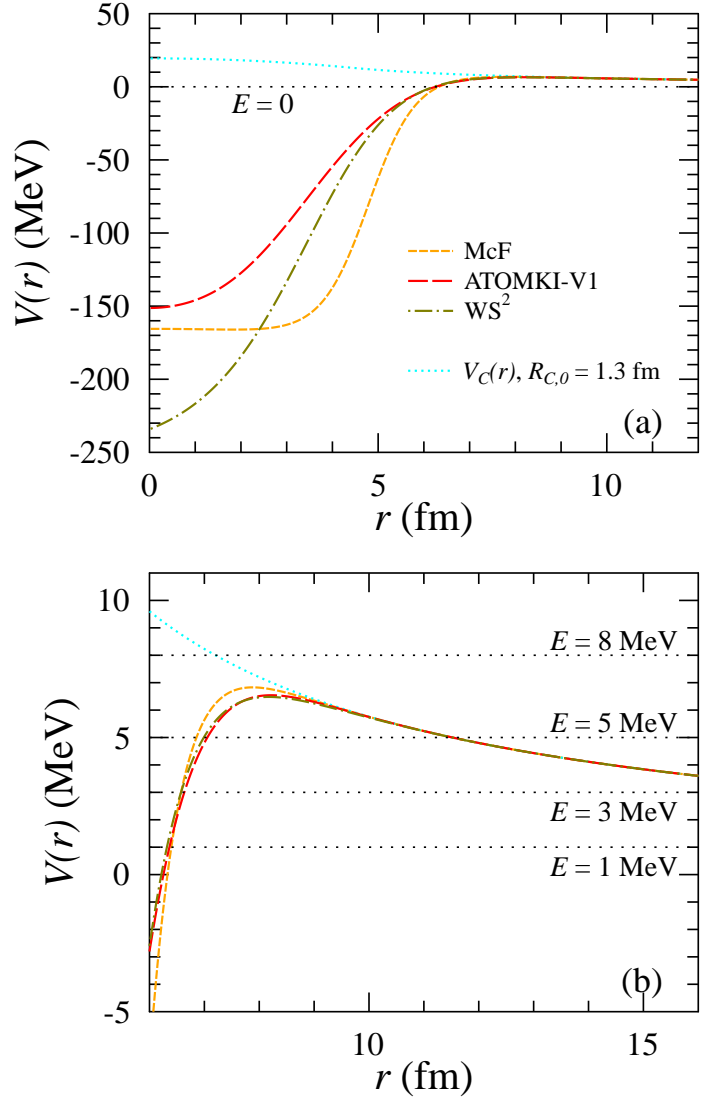}}
\caption{
  \label{fig:pot}
  Real part of the potential $V(r) = V_N(r) + V_C(r)$. The nuclear potentials
  $V_N(r)$ under study are the McFadden/Satchler potential
  \cite{McFadden_NPA1966_aomp}, the ATOMKI-V1 potential
  \cite{Mohr_ADNDT2013_atomki-v1}, and the squared Woods-Saxon potential by
  Koyuncu and Soylu \cite{Koyuncu_IJMPE2017}; the potentials are explained in
  Sec.~\ref{sec:pot}. The repulsive $V_C(r)$ is shown by a lightblue dotted
  line. The upper part (a) shows the overall attractive $V(r)$; the lower part
  (b) shows the resulting barrier around $r \approx 8$ fm. In addition, four
  energies are investigated in more detail; these are marked by dotted
  horizontal lines (see discussion in Sec.~\ref{sec:disc}).  }
\end{figure}

The real parts of the three chosen potentials are shown in
Fig.~\ref{fig:pot}. There is a significantly different behavior in the nuclear
interior, see part (a) of Fig.~\ref{fig:pot}. However, the three potentials
under study show quite similar Coulomb barriers with a slightly higher barrier
for the McF potential, see part (b) of Fig.~\ref{fig:pot}. It is interesting
to note that the \WSsq\ potential of \KS\ was fitted to a double-folding
potential. Although the present ATOMKI-V1 double-folding potential may be
slightly different from the \KS\ double-folding (e.g., because of the chosen
density parametrizations or because of a slightly different normalization),
the resulting barriers for the \WSsq\ potential from \KS\ and the ATOMKI-V1
double-folding potential are almost identical. Consequently, at least in the
pBTM very similar cross sections should result (in conflict with the
conclusion of \KS ).

\subsection{Results}
\label{sec:res}
\begin{figure}[th]
\centerline{\includegraphics[width=0.75\columnwidth]{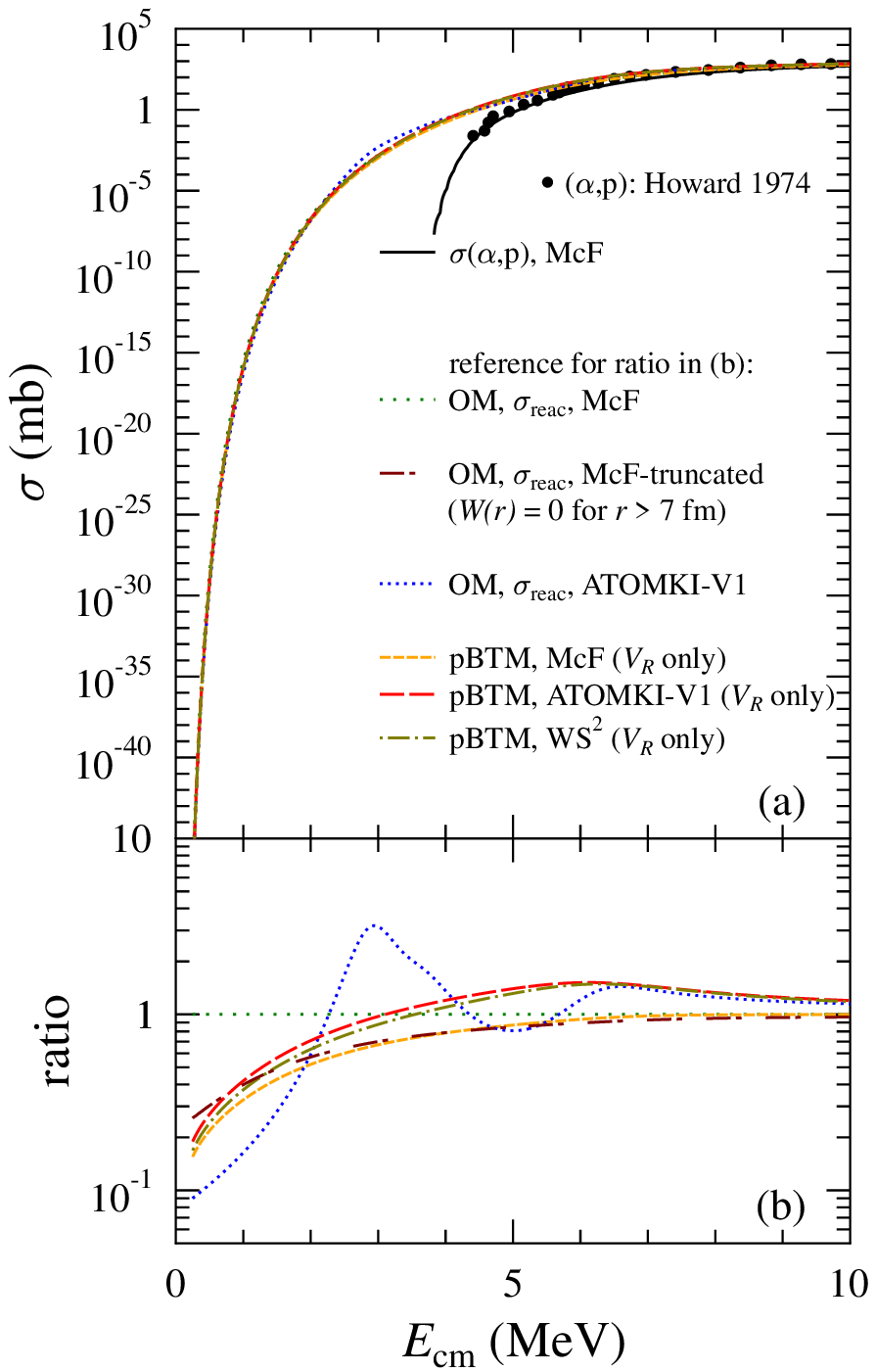}}
\caption{
  \label{fig:sigtot}
  Cross section of \al\ + \canull\ as a function of energy: The upper part (a)
  shows the cross sections \stot\ and the \canull \rap \sciii\ cross section
  from the McF potential in the statistical model; the lower part (b) shows
  the ratios of the calculated cross sections normalized to the reference
  cross section (OM, McF potential). Experimental data are taken from Howard
  {\it et al.}\ \cite{Howard_APJ1974_a-X}. Further discussion see text.
}
\end{figure}

The calculated cross sections cover many orders of magnitude from almost 1
barn at the highest energies around 10 MeV down to tiny cross sections of the
order of $10^{-50}$ barn at the lowest energies under study. Thus, for better
visualization, Fig.~\ref{fig:sigtot} shows the calculated cross sections in
the upper part (a) and normalized cross sections in the lower part (b). For
normalization, the OM calculation using the McF potential was used as a
reference.

For comparison with experiment, the data of Howard {\it et
  al.}\ \cite{Howard_APJ1974_a-X} can be used. This experiment has measured
the \sciii\ production cross section which results from the \canull \ran
\tiiii\ and \canull \rap \sciii\ reactions. In the energy range under study,
the \ran\ channel is closed because of the strongly negative $Q$-value of $Q =
-11.2$ MeV, and thus the \sciii\ production results from the \rap\ channel
only. Also the \rap\ channel has a negative $Q$-value of $Q = -3.5$ MeV. At
lower energies the \rag\ capture channel is dominating in the
calculations. Above about 6 MeV, the \rap\ cross section exceeds the
\rag\ cross section by far, and thus the \rap\ cross section approaches the
total cross section \stot ; other particle channels like (\al ,2p) or (\al
,np) are also closed. Under these conditions the SM calculation for the
\rap\ cross section practically depends only on the chosen \al -nucleus
potential. The SM calculation with the McF potential provides a good
description of the experimental \rap\ cross sections (full black line in
Fig.~\ref{fig:sigtot}), and it is known that the McF potential is able to
reproduce \al -induced cross sections in this mass range very well
\cite{Mohr_EPJA2015_A20-50,Mohr_PRC2018_38ar}. From the \rap\ threshold to
about 6 MeV, the calculated \rap\ cross section results from the total
reaction cross section \stot\ (depending on the chosen \al -nucleus potential)
and from the branching towards the proton channel which is suppressed by the
small proton transmission close above threshold (depending on the chosen
proton-nucleus potential, $\gamma$-ray strength, and level density). The
excellent agreement betwen the SM calculation and the experimental data down
to the lowest data points around 4.5 MeV can be considered as further
confirmation of the \al -nucleus potentials under study, but similar cross
sections could also be obtained from different combinations of the various
ingredients of the SM calculations (for a discussion of the different
ingredients of the SM calculations, see
\cite{Mohr_PRC2017_64zn}). Unfortunately, there are no experimental data which
can constrain the \al -nucleus potentials below the \rap\ threshold at 3.5
MeV.

The ATOMKI-V1 potential predicts slightly larger cross sections than the McF
potential for energies above 5 MeV in the OM. At very low energies, the
ATOMKI-V1 cross sections in the OM are far below the McF cross sections, and
there are surprisingly large ATOMKI-V1 cross sections around 3 MeV. This
somewhat unexpected energy dependence will be analyzed below (see
Sec.~\ref{sec:disc}).

Sec.~\ref{sec:disc} will also explain the truncated McF potential where the
imaginary part $W(r)$ was set to zero for radii $r > 7$ fm. This truncated McF
potential leads to cross sections in the OM which are very close to the
results from the pBTM (using the real part of the potential only).

For the pBTM calculations, the real parts of the McF and ATOMKI-V1 potentials
were used. In addition, the results from the \WSsq\ potential of \KS\ are
shown. In general, at higher energies above the barrier of about 6.5 MeV, for
each potential there is excellent agreement between the full OM calculation
(using the complex nuclear potential) and the simpler pBTM calculation (using
only the real part of the nuclear potential). At lower energies, the pBTM
cross sections are generally lower than the full OM calculations. All pBTM
calculations show a very similar and smooth energy dependence, and the
deviations for the three potentials under study remain within about a factor
of two over the whole energy range in Fig.~\ref{fig:sigtot}. As expected, the
pBTM calculations from the ATOMKI-V1 double-folding potential and the
\WSsq\ potential of \KS\ remain very close because the barriers are almost
identical for these potentials. This result is in clear contradiction to the
conclusion of \KS\ where the cross section of the double-folding potential in
the WKB approximation exceeds the \WSsq\ potential by many orders of magnitude
at low energies.

\begin{figure}[th]
\centerline{\includegraphics[width=0.70\columnwidth]{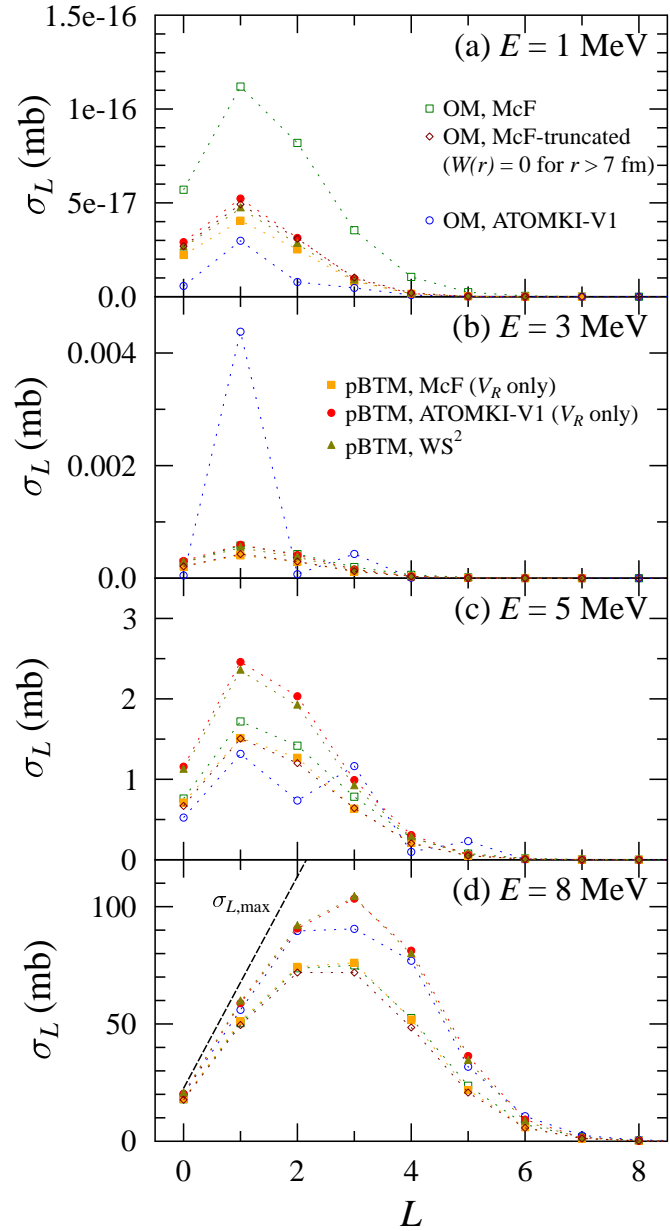}}
\caption{
  \label{fig:sigl}
  Partial cross sections $\sigma_L$ for \al\ + \canull\ for four selected
  energies between 1 MeV and 8 MeV. The dashed line at 8 MeV indicates
  $\sigma_{L,{\rm{max}}}$ from Eq.~(\ref{eq:siglmax}). The thin dotted lines
  connect the data points to guide the eye. Further discussion see
  text.
}
\end{figure}

For a deeper understanding of the results, four energies have been selected
which are 1, 3, 5, and 8 MeV. At these energies one can see interesting
properties of the calculated excitation functions. Very low cross sections
from the ATOMKI-V1 OM calculation are found at 1 MeV, and there is a
significant difference between the OM and pBTM calculations using the McF
potential; very high cross sections from the ATOMKI-V1 OM calculation are
found at 3 MeV; again low ATOMKI-V1 cross sections appear at 5 MeV; and almost
identical cross sections from all potentials and models can be seen at 8 MeV,
i.e.\ above the Coulomb barrier. For these four selected energies the partial
$\sigma_L$ cross sections are shown in Fig.~\ref{fig:sigl} and discussed in
the following Sec.~\ref{sec:disc}.

\section{Discussion}
\label{sec:disc}
Advantages and disadvantages of the pBTM and OM model have already been
discussed above in Sec.~\ref{sec:comp}. These advantages and disadvantages
are also reflected in the calculated partial
$\sigma_L$ at the selected energies of 1, 3, 5, and 8 MeV (see
Fig.~\ref{fig:sigl}). Only a relatively small number of partial waves
contributes to the sum in Eq.~(\ref{eq:BTM}) or Eq.~(\ref{eq:OM}). Note that a
linear scale was chosen in Fig.~\ref{fig:sigl} for best visualization of the
relevant $\sigma_L$ from the different calculations. Even significant
variations from different potentials for large $L$ do not affect the
calculated \stot\ in Eqs.~(\ref{eq:BTM}) and (\ref{eq:OM}) because $\sigma_L$
for large angular momenta $L$ are much smaller than the dominating $\sigma_L$
for small $L$.

At energies above the barrier, e.g.\ at 8 MeV in part (d) of
Fig.~\ref{fig:sigl}, the transmission $T_L$ for small angular momenta $L$ in
the pBTM approach unity, leading to the maximum contribution
$\sigma_{L,{\rm{max}}}$ of these partial waves:
\begin{equation}
  \sigma_{L,{\rm{max}}} = \frac{\pi}{k^2}  \, \times \, (2L+1) \quad \quad .
  \label{eq:siglmax}
\end{equation}
This maximum contribution is indicated as a dashed line in
Fig.~\ref{fig:sigl}, part (d). A similar behavior is found for the OM where
the partial waves with small angular momenta $L$ enter the region with strong
imaginary part, leading to full absorption of these partial waves (reflexion
coefficients $\eta_L \approx 0$). Thus, also in the OM the $\sigma_L$ approach
their maximum value in Eq.~(\ref{eq:siglmax}). This finding is independent of
details of the chosen potentials, leading to similar $\sigma_L$ from the OM
and from the pBTM and for all potentials. Indeed, at 8 MeV the calculated
cross sections are within a relatively narrow range of 300 mb $\lesssim$
\stot\ $\lesssim$ 400 mb, i.e., the variations are of the order of 30\%
only. A complete discussion on the general behavior of $\sigma_L$ at different
energies was given in previous work \cite{Mohr_PRC2013_ce140a}.

Slightly below the Coulomb barrier at 5 MeV, the role of the barrier becomes
relevant. The slightly higher barrier from the McF potential leads to smaller
cross sections than ATOMKI-V1 and \WSsq\ in the pBTM. However, \stot\ from the
ATOMKI-V1 potential in the OM is smaller than all other calculations. This is
related to an odd-even staggering in the $\sigma_L$, see part (c) of
Fig.~\ref{fig:sigl}, which is a typical feature for imaginary potentials with
a surface-only shape and has also been observed in previous work
\cite{Mohr_PRC2016_a-n}.

At 3 MeV this odd-even staggering becomes most pronounced, see part (b) of
Fig.~\ref{fig:sigl}. Only one partial wave with $L=1$ is responsible for the
enhancement of the ATOMKI-V1 potential in the OM calculation. The other
calculations show a smoother energy dependence of the excitation functions in
Fig.~\ref{fig:sigtot} and also a smoother $L$ dependence of the $\sigma_L$ in
Fig.~\ref{fig:sigl}. Similar to the 5 MeV result, the cross sections in the
pBTM are larger for the ATOMKI-V1 and \WSsq\ potentials than for the McF
potential.

A dramatic effect is also seen in the OM calculation for the ATOMKI-V1
potential at very low energies. Contrary to the relatively high \stot\ at 3
MeV, now the ATOMKI-V1 potential predicts a very low \stot\ about a factor of
six below the OM calculation with the McF potential.

At the lowest energy of 1 MeV in part (a) of Fig.~\ref{fig:sigl}, the
difference between the OM and the pBTM for the same McF potential increases to
more than a factor of about 3. This difference results from the properties of
the different models (pBTM vs.\ OM). In the pBTM, the incoming \al -particle
has to tunnel through the Coulomb barrier, leading to small cross sections in
the pBTM. The same transmission is also calculated in the OM. But, in the OM,
some additional absorption may already happen at larger radii without complete
tunneling because of the tail of the imaginary potential towards larger
radii. Cutting the tail of the imaginary part of the McF potential in the OM
calculation at $r = 7$ fm reduces the OM cross section of \stot\ by a factor
of three and brings the OM result very close to the pBTM
calculation. Interestingly, this finding holds for the whole energy range
under study in this work: both, the total cross sections in
Fig.~\ref{fig:sigtot} and the partial cross sections $\sigma_L$ at all
energies in Fig.~\ref{fig:sigl}, are very similar from the OM and the McF
potential with the truncated imaginary part on the one hand and from the pBTM
(which uses only the real part of the McF potential) on the other hand. Note
that the depth of the imaginary part of the McF potential at 7 fm is only
$-0.35$ MeV (or about 1.4\% of its central value of $-25$ MeV); i.e., the tiny
tail of the imaginary part at large radii is mainly responsible for the
calculated cross section in the OM at very low energies. Or, in other words,
according to the OM calculation, the dominating contribution to the reaction
cross section results from radii $r > 7$ fm which is at the nuclear surface
and even beyond. This finding clearly illustrates the sensitivity of the OM
calculations to the chosen parametrization of the imaginary potential at large
radii.

It is well-known that the cross sections of many transfer and capture
reactions for light nuclei at low energies are essentially defined by the
asymptotic properties of the wave functions at large radii, see e.g.\ the
broad discussion of the $^{16}$O\rpg $^{17}$F reaction in literature
\cite{Rolfs_NPA1973_p-g,Chow_CJP1975_16o_p-g,Morlock_PRL1997_16o_p-g,Gagliardi_PRC1999_16o_p-g,Iliadis_PRC2008_16o_p-g,Blokhintsev_PRC2018_16o_p-p}. The
concept of the asymptotic normalization coefficient (ANC) has been developed,
and it has been shown that various quantities in nuclear reactions are related
to the ANC \cite{Mukhamedzhanov_PRC1999_ANC}. However, the observed
sensitivity of the OM reaction cross sections to the tail of the imaginary
potential at large radii differs somewhat from the ANC concept because this
sensitivity is also relevant for calculations in the statistical model and
thus applies to compound reactions for intermediate mass and heavy nuclei. The
relevance of the \al -nucleus potential has been noticed in many recent
studies of \al -induced reaction cross sections. The important role of the
imaginary part is obvious from Eqs.~(\ref{eq:OM}) and (\ref{eq:SL}), and
modifications of the imaginary strength have been suggested to reproduce
experimental data at low energies. However, the importance of the tail of the
imaginary potential at sub-Coulomb energies was not pointed out, see
e.g.\ recent work
\cite{Sauerwein_PRC2011_141pr,Halasz_PRC2012_130ba_132ba,Palumbo_PRC2012_a-n,Palumbo_PRC2012_a-a,Rauscher_PRC2012_169tm,Kiss_PRC2012_127i,Kiss_PRC2013_113in,Netterdon_NPA2013_168yb,Glorius_PRC2014_165ho_166er,Yalcin_PRC2015_107ag,Netterdon_PRC2015_112sn,Simon_PRC2015_ni_a-X,Quinn_PRC2015_a-X,Halasz_PRC2016_124xe,Ornelas_PRC2016_64zn,Scholz_PLB2016_aomp,Korkulu_PRC2018_sb_a-X,Kiss_PRC2018_115in}
on \al -induced reactions at low energies from the last decade.

The observed sensitivity of the low-energy cross section in the OM calculation
to details of the imaginary part leads to the open question whether any OM
calculation is able to predict reliably cross sections far below the Coulomb
barrier. Because these low-energy cross sections are tiny, an experimental
verification of any OM prediction seems to be very difficult. This sensitivity
may also -- at least partly -- be responsible for the large range of predicted
\rag\ cross sections for heavy target nuclei, as e.g.\ found in the pioneering
work of Somorjai {\it et al.}\ for the $^{144}$Sm\rag $^{148}$Gd reaction
\cite{Somorjai_AaA1998_sm144ag}. In contrast to the OM calculations, the pBTM
results show a much smoother energy dependence for all potentials under
study. This finding further confirms that the imaginary part of the OM
potential is a very delicate ingredient for the calculation of low-energy
cross sections.

For completeness it has to be pointed out that the cross sections for
\canull\ + \al\ at the energy of 1 MeV correspond to a Gamow window at $T_9
\approx 0.3$ (where $T_9$ is the stellar temperature in $10^9$ K). Because of
the very small cross section at these low energies, the astrophysical reaction
rate \Nsv\ becomes practically negligible. Consequently, the cross sections at
even lower energies below 1 MeV are not very relevant. Furthermore, at these
low energies the \canull\ + \al\ cross section is dominated by the \canull
\rag \tiiv\ capture channel because the \rap\ and \ran\ channels are
closed. It is well-known that the \canull \rag \tiiv\ cross section at low
energies is governed by individual resonances
\cite{Robertson_PRC2012_ca40ag,Schmidt_PRC2013_ca40ag}, and calculations in
the OM or pBTM can only provide average cross sections over a broader energy
interval.

Finally, almost identical cross sections are found in the pBTM from the
ATOMKI-V1 double-folding potential and from the \WSsq\ potential of \KS . This
finding is expected from the similar shape of the Coulomb barrier for these
potentials (as shown in Fig.~\ref{fig:pot}), but it is also in clear
contradiction to the conclusion of \KS . From a discussion with \KS\ during
the review process of this paper it became clear that the unexpected huge
cross sections of \KS\ for their double-folding potential are related to the
calculation of the turning points in the WKB approximation, and thus their
strong conclusion ``DF potential $\ldots$ has failed to produce \al\ +
\canull\ cross sections at low energies'' should be somewhat weakened to ``the
DF potential did not lead to reasonable results in the WKB calculations''. The
present study clearly confirms that double-folding potentials can be applied
to \canull\ + \al\ at low energies within the pBTM and the OM.

\section{Summary and conclusions}
\label{sec:summ}
The total reaction cross section \stot\ for \canull\ + \al\ was calculated at
low energies around and below the Coulomb barrier using two different models
and three different \al -nucleus potentials. The excitation functions in the
barrier transmission model (using a real nuclear potential) show a smooth
energy dependence whereas the excitation functions in the optical model (using
a complex nuclear potential) show a significant dependence on the chosen
parametrization of the imaginary part of the potential, in particular in the
case of the ATOMKI-V1 potential with an imaginary part of Woods-Saxon surface
type.

In general, towards lower energies the cross sections from the barrier
transmission model become smaller than the optical model cross sections. This
result can be explained by additional absorption contributions in the tail of
the imaginary potential for large radii in the optical model (i.e., in the
nuclear surface region, in the Coulomb barrier, and even beyond). As this tail
of the imaginary potential is not very well constrained by any experimental
data, this leads to significant uncertainties in the prediction of total
reaction cross sections at very low energies in all optical model calculations
and also in the statistical model which is based on the total reaction cross
section \stot\ from the optical model. These uncertainties are very important
for the calculation of astrophysical reaction rates for \al -induced reactions
which are mainly sensitive to the cross sections far below the Coulomb barrier
in the so-called Gamow window. At higher energies above the Coulomb barrier
the cross sections from the different models and from the different \al
-nucleus potentials converge nicely.

Finally, in line with many previous publications, but in contradiction to the
conclusion of a recent study within the WKB approximation by Koyuncu and Soylu
\cite{Koyuncu_IJMPE2017}, it is found that double-folding potentials together
with the pure barrier transmission model and the optical model can be well
applied for \canull\ + \al\ at low energies.

\section*{Acknowledgements}
I thank Zs.\ F\"ul\"op, Gy.\ Gy\"urky, G.\ G.\ Kiss, and F.\ Koyuncu for
encouraging discussions. This work was supported by NKFIH (K108459 and
K120666).

\bibliographystyle{ws-ijmpe}
\bibliography{pm}

\end{document}